\documentclass{moriond}
\usepackage{indentfirst} % indent the first paragraph of a section

\def\be{\begin{equation}}
\def\ee{\end{equation}}
\def\bea{\begin{eqnarray}}
\def\eea{\end{eqnarray}}

\newcommand{\Photo}{}

\begin{document}
\vspace*{4cm}
\title{Building on SVOM : the CATCH satellite constellation	for transient astronomy}

\author{ Stéphane Schanne}

\address{CEA Paris-Saclay, DRF/IRFU/DAp-AIM, 91191 Gif sur Yvette, France \\
{\rm on behalf of the CATCH consortium (IAMC, IHEP, NAOC, CNES, CEA, IRAP, APC)}
}

\maketitle\abstracts{
%compress by half:
Building on the success of the French–Chinese SVOM mission, our consortium proposes CATCH (Chasing All Transients Constellation of Hunters), targeting Gamma-Ray Bursts and X-ray transients. Its first step, CATCH-PM (Precursor Mission), consists of 3 satellites: (i) the Trigger Scout Satellite (TSS), providing real-time detection and localization of transients from soft X-rays to $\gamma$-rays; (ii) the X-ray Hunter Satellite (XHS), performing rapid X-ray afterglow follow-up and refined localization with Wolter-I optics; and (iii) the Infrared Hunter Satellite (IHS), enabling precise localization and follow-up in the visible and near-infrared, including obscured and high-redshift transients.
CATCH-PM is both a technological demonstrator and a standalone time-domain astrophysics mission. In addition to Scout triggers, it will accept numerous external Targets of Opportunity and conduct a broad observing program. By reusing available spare hardware, a launch within 5 years is feasible. Operating alongside SVOM and EinsteinProbe it would significantly enhance the overall scientific return.
}

\section{Introduction and scientific domain}\label{subsec:intro-science}

%compress by half:
CATCH (\emph{Chasing All Transients Constellation of Hunters}) is a proposed space mission\cite{catch-old} consisting of a constellation of satellites dedicated to high-energy transient astronomy.

These transients include gamma-ray bursts (GRBs): (i) long GRBs, linked to the collapse of massive stars, and (ii) short GRBs, produced by binary neutron star mergers. Both generate relativistic jets responsible for prompt emission (from fractions of a second to minutes, mainly in hard X-ray/$\gamma$-rays) and afterglow emission observed from X-rays to radio over hours to days. Other sources include (iii) stellar-mass X-ray binaries, where accretion onto a neutron star or black hole produces high-energy emission, and (iv) supermassive black holes, such as active galactic nuclei and blazars, which generate transients through particle acceleration.

The study of high-energy transients probes extreme physics (strong gravity, intense electromagnetic fields, and ultra-dense matter) and is often connected to gravitational waves and high-energy neutrinos. These phenomena provide key insights into compact objects, central engines, and non-thermal processes beyond the reach of laboratory experiments.

\section{Current and future hard X-ray transient real-time missions}\label{subsec:current-missions}

%cleanup and slighly reduce in size:
Current missions dedicated to real-time detection, localization, and follow-up of hard X-ray transients include: (i) Swift, a highly successful NASA–ASI mission operational since 2004, which suspended scientific operations in January 2026 pending an orbital reboost; (ii) EinsteinProbe, a mission of CAS (Chinese Academy of Sciences) with ESA, MPE and CNES participation, launched in January 2024; and (iii) SVOM (Space-based multi-band Variable Objects Monitor\cite{svom}), a mission of CNSA (Chinese National Space Administration), CAS and CNES with contributions from French laboratories (CEA, CNRS), Universities and MPE, launched in June 2024. The latter two missions are expected to operate for at least 5 possibly up to 10 years.
Looking ahead, the Theseus mission, an ESA M7 candidate, could take over a major role in real-time high-energy transient studies, if selected in summer 2026, with a launch no earlier than 2037.

In the meantime, to bridge this gap, CATCH is proposed as a continuation of SVOM by the same Chinese–French consortium involving CNSA, CAS and CNES with French laboratories. The mission is planned in two stages: CATCH-PM (Precursor Mission\cite{catch-new}) followed by CATCH-FOM (Full-Operational Mission). Here we focus on CATCH-PM, targeting a launch within five years, and aiming joint operations with SVOM and EinsteinProbe to maximize scientific return.

In the SVOM concept, primarily dedicated to GRBs, a single satellite platform carries four instruments: two for prompt emission detection (GRM and ECLAIRs) and two for afterglow follow-up (MXT and VT). GRM is a non-imaging detector for $\gamma$-ray spectroscopy. ECLAIRs is a wide-field (2 sr) coded-mask hard X-ray imager equipped with an onboard trigger\cite{trigger} system that detects and localizes transients to better than 11 arcmin (at the detection limit), sends real-time alerts via a dedicated VHF network, and initiates spacecraft slews. Within $<$2 minutes, the narrow-field instruments, MXT (an X-ray telescope using lightweight micropore optics) and VT (a visible telescope operating simultaneously in blue and red bands), perform afterglow observations and refined localization, enabling ground-based follow-up, including spectroscopy for redshift determination and complete event characterization.

\section{The CATCH-PM concept}\label{subsec:catch-pm}

%cleanup and slighly reduce in size:

\begin{figure}
	\centering
	\includegraphics[width=1.0\textwidth, angle=0]{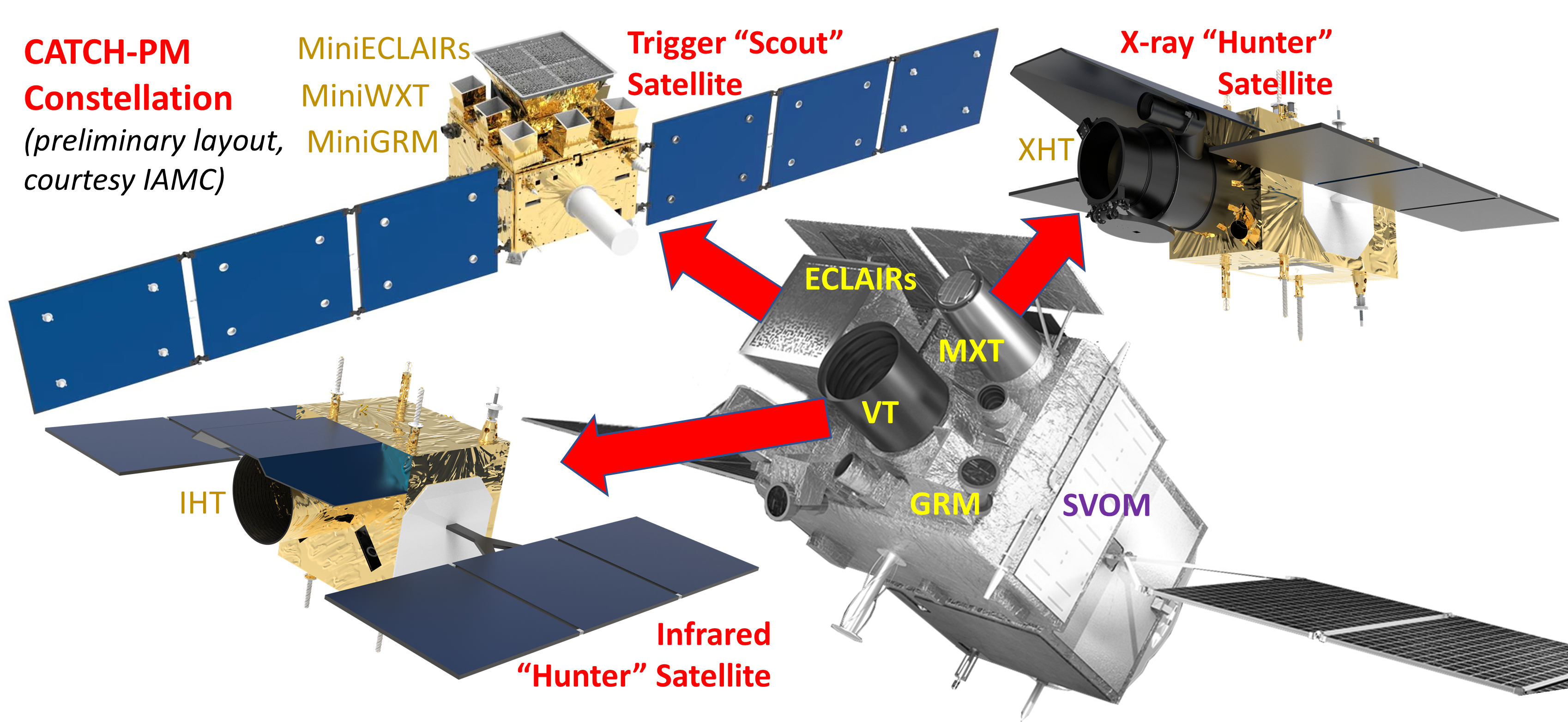}
	%\caption{The CATCH-PM constellation with instruments distributed over 3 satellites, compared to the monolithic satellite SVOM carrying 4 instruments.}
	\caption{CATCH-PM constellation based on 3 satellites, compared to the monolithic satellite SVOM.}
	\label{fig_catch_layout_vs_svom}
\end{figure}

The concept of the mission CATCH-PM (Fig. \ref{fig_catch_layout_vs_svom}) is to distribute the instruments across a constellation of three satellites: the Trigger Scout Satellite (TSS), the X-ray Hunter Satellite (XHS), and the Infrared Hunter Satellite (IHS). Each satellite is in the 300–500 kg class and is based on platforms developed by IAMC (Innovation Academy for Micro-satellites of the CAS in Shanghai), which previously built the SVOM and EinsteinProbe satellites. 
%CATCH-PM is conceived as a technological demonstrator, reusing flight spare parts with a target launch date within about five years.

CATCH-PM will implement the “satellite train” concept, with the spacecrafts flying close to each other (separation $<$1000 km). The Scout will trigger on the appearance of a transient and send the alert via an inter-satellite link to the Hunters, which will then slew rapidly (within $<$5 minutes) to perform follow-up observations, while observations of the Scout continue.
The key concept of the “satellite train” is therefore to decouple the trigger from the follow-up:

(a) {\bf The Scout satellite} will not be required to perform a slew, allowing continuous observation of the transient without interruption at critical phases, such as during the prompt or prompt-to-afterglow transition. This maximizes photon collection and improves the signal-to-noise ratio and localization. The Scout will follow an anti-solar pointing strategy (similar to SVOM), allowing rapid ground-based follow-up of the burst detected during night. It will mostly avoid the Galactic plane, where bright sources reduce trigger efficiency. It will however readjust its pointing each orbit to reduce the Earth fraction in the large field of view of the trigger instruments (to reach a fraction $>$80\% free of Earth). In contrast for SVOM, without such maneuvers, the Earth-free portion of the ECLAIRs 2-sr field of view is 65\% on average. These Earth-avoidance maneuvers will gain observing time and periodically bring the Galactic plane into the field of view (though reducing trigger efficiency) and enable regular monitoring of Galactic transient sources, enhancing the non-GRB observing program.

(b) {\bf The Hunter satellites} will have significantly greater flexibility in this “train” concept compared to a single-platform configuration. Each Hunter will independently conduct follow-up observations of the transient, for as long as it remains visible for its telescope. And when waiting for a burst to follow-up, rather than being constrained to observe mostly at high Galactic latitudes (as in SVOM, whose pointing strategy avoids the Galactic plane and Sco X-1 to optimize trigger efficiency), the CATCH Hunters will have much greater freedom in their observing programs. They will be able to perform regular and detailed observations of Galactic plane sources, as well as respond to a large number of external triggers uploaded rapidly from the ground as Targets of Opportunity (ToO) observations.

%\section{The CATCH-PM orbit and system}\label{subsec:catch-pm-system}

%cleanup and slighly reduce in size:

The CATCH-PM satellites will be launched together by the same launcher into low Earth orbit (about 650 km altitude), from near Sanya on the Hainan island, located at 19$^\circ$ latitude. Therefore, the orbit inclination will be lower than for SVOM (launched from Xichang at about 30$^\circ$ latitude), reducing the time spent in the SAA (South Atlantic Anomaly of the Earth magnetic field) and in the outer radiation belts, where the particle background is much enhanced. The time fraction for data collection by the trigger instrument is thus expected to be above 75\% (while it is about 60\% for ECLAIRs at current solar maximum).

The CATCH-PM will also reuse the SVOM VHF network for alert and follow-up real-time data downlink to ground (while requiring fewer receiver stations thanks to the lower orbit inclination). ToOs will be uploaded via Beidou, as for SVOM. Other solutions, as well as the inter-satellite link, are under study. The full data will be downloaded with delay (several hours) via X-band. The CATCH-PM will be inserted into the existing SVOM science ground segment (FSC and CSC will be reused), as well as into the mission control at NSSC in Beijing.

\section{The CATCH-PM satellites and instruments}\label{subsec:catch-pm-instruments}

%cleanup and slighly reduce in size:

{\bf The Trigger Scout Satellite (TSS)} is dedicated to the detection of X-ray transients and GRB prompt emission, and will for the first time enable simultaneous sky monitoring in soft and hard X-rays. It will perform long uninterrupted observations of transients, rapidly alert the Hunters and the ground, and allow a rich non-GRB science program.
The Scout will carry two triggering instruments, the French MiniECLAIRs and the Chinese MiniWXT (as well as a GRM- or GECAM-like spectrometer to characterize prompt emission up to several MeV):

(a) The MiniECLAIRs, proposed to be provided by France, imaging the sky from 4 to 120 keV, will reuse 50 ECLAIRs flight-spare detector modules and associated electronics. This enables a detection plane of 250 cm$^2$ active area (1/4 of ECLAIRs), located $\sim$37 cm below a coded mask, identical in geometry to that of ECLAIRs, providing a 2 sr total field of view and a fully coded field of view of 0.8 sr (compared to 0.18 sr for ECLAIRs), with localization accuracy better than $\sim$15 arcmin at the detection limit. The onboard software (developed at CEA), including the Count-Rate Trigger (CRT) and Image Trigger (IMT) algorithms, will be adapted to the MiniECLAIRs geometry and onboard computer. Based on simulations using the ECLAIRs GRB sample scaled to MiniECLAIRs, including orbital and Earth-free visibility  conditions, the expected detection rate is about 45–50 GRBs per year (vs. 60 for ECLAIRs).

(b) The MiniWXT, built in China, imaging in the 0.5–4 keV band, consists of $\sim$8 mini-cameras using micropore optics coupled to CMOS detectors, inherited from the EinsteinProbe WXT instrument, with reduced focal distance in order to enlarge the field of view  ($>$0.5 sr), placed inside the fully coded field of view of MiniECLAIRs and achieving typical localizations $<$3 arcmin.
The Scout platform computer will cross-match triggers from MiniWXT and MiniECLAIRs, reducing the false trigger rate, before requesting the slew of the Hunter satellites.

{\bf The X-ray Hunter Satellite (XHS)} is dedicated to X-ray follow-up of transients using a powerful 0.3–10 keV telescope, enabling faint source detection, fine localization ($<$10 arcsec), and long-term follow-up (typically tens of hours). XHS receives alerts from the Scout or rapid ToO requests from the ground and transmits real-time results via VHF and inter-satellite link. Between alerts, it performs a broad observatory program, including Galactic plane surveys, with high flexibility and access to $\sim$80\% of the sky thanks to a large sun-shield. The telescope is similar to the EinsteinProbe FXT, reusing spare Wolter-I optics available in China, with an effective area of $\sim$250 cm$^2$ (2$\times$ Swift/XRT and 10$\times$ SVOM/MXT). The camera is proposed to be provided by France, reusing the flight-spare model of the SVOM/MXT camera available at CEA. The resulting field of view of 40$\times$40 arcmin matches to the trigger localization accuracy.

{\bf The Infrared Hunter Satellite (IHS)} is dedicated to simultaneous near-infrared and visible follow-up of transients and represents a major advance. It will enable detection and lightcurve monitoring of highly redshifted GRBs (burst at z$>$6 are redshifted outside of the SVOM/VT band) and dark GRBs (which represent $\sim$25\% of the events followed by SVOM/VT). Its precise localization accuracy ($<$1 arcsec) will enable ground-based spectroscopic follow-up for redshift determination. IHS receives alerts from the Scout and position refinements from XHS or rapid ToO requests from the ground and sends real-time results via VHF and inter-satellite link. Between alerts, it also conducts a very rich and flexible observatory program. It reuses the Chinese SVOM/VT qualification model of 44 cm diameter and 26$\times$26 arcmin field of view. A beam splitter enables simultaneous infrared (1000–1700 nm) and visible observations (blue: 400–650 nm or red: 650–1000 nm selected via filter wheel). The infrared detector, built in China, could be characterized in France at CEA's infrared test facility.

\section{Conclusions}\label{subsec:catch-conclusions}

%cleanup and slighly reduce in size:

The CATCH-PM (Precursor Mission of the CATCH constellation), a first constellation demonstrator, has a very strong science case in real-time transient astronomy and will deliver first-class results through its three satellites: (i) the Trigger Scout Satellite, providing simultaneous X-ray to $\gamma$-ray triggering and transient localization; (ii) the X-ray Hunter Satellite, equipped with a powerful and versatile X-ray telescope; and (iii) the Infrared Hunter Satellite, which represents a major advance thanks to its space-based infrared telescope.

CATCH-PM, currently studied by a joint French–Chinese team, continues the successful technical and scientific collaboration initiated with SVOM. It builds on the expertise of IAMC in minisatellites, demonstrated by the SVOM and EinsteinProbe missions. The project maintains a balanced partnership, in the spirit of SVOM, and reuses key elements, including the VHF network and ground segment. By relying on space-qualified spare components, it can be implemented in a fast and cost-effective manner. With a launch targeted around 2030, joint operations with SVOM and EinsteinProbe will further enhance the overall scientific return.

\section*{References}
%\bibliography{moriond}

%%% manually generated bibliography

\end{document}